# Deformation and failure maps for PMMA in uniaxial tension

F. Van Loock and N.A. Fleck

Engineering Department, Cambridge University, Trumpington Street,
Cambridge CB2 1PZ, United Kingdom

April 2018

**Abstract**

Uniaxial tensile tests are performed on a polymethyl methacrylate (PMMA) grade over a range of temperatures near the glass transition and over two decades of strain rate. Deformation maps are constructed for Young's modulus, flow strength, and failure strain as a function of temperature for selected strain rates. The glassy, glass transition and rubbery regimes are identified, and constitutive relations are calibrated for the modulus, flow strength and failure strain within each regime.

## 1. Introduction

The linear, amorphous version of the polymer polymethyl methacrylate (PMMA) is employed in a variety of technical products ranging from transparent windshields to nanocellular polymeric foams [1,2]. Commonly, these components are formed in the hot state at a temperature close to the glass transition temperature $T_g$. However, measurements of the tensile stress-strain response of PMMA near $T_g$ are limited. Our aim is to identify the regimes of deformation and fracture for a commercial PMMA grade near the glass transition. Existing constitutive models are calibrated and an emphasis is placed on the magnitude of the failure strain as a function of temperature and strain rate.

Gilbert et al. [3] have identified four constitutive regimes for a linear, amorphous polymer under uniaxial tension. These regimes are the (1) glassy, (2) glass transition, (3) rubbery, and (4) viscous flow regimes in sequence of increasing temperature T or of decreasing strain rate $\dot{\varepsilon}$, see Figure 1a. The typical stress versus strain response associated with each regime is illustrated in Figure 2a (nominal tensile stress S versus nominal tensile strain e) and b (true tensile stress σ versus true (logarithmic) tensile strain ε). The deformation mechanism map of Gilbert et al. [3] made use of data in the literature for various PMMA grades as insufficient comprehensive data exist for a single grade. A series of simplified constitutive laws were fitted to the experimental modulus data. In a subsequent study, Ahmad et al. [4] reported similar plots for the rate and temperature dependent flow strength $\sigma_y$ by an analogous approach, see Figure 1b.

Few data exist in the literature on the true tensile failure strain $\varepsilon_f$ of PMMA as a function of temperature and strain rate. A possible trend is depicted in Figure 1c, as based on Cheng et al. [5], Smith [6] and Vinogradov et al. [7]. Cheng et al. [5] measured $\varepsilon_f$ in the glassy regime, Smith [6] elucidated the effect of rate and temperature on $\varepsilon_f$ for cross-linked elastomers near the glass transition temperature, and Vinogradov et al. [7] investigated the effect of strain rate $\dot{\varepsilon}$ on $\varepsilon_f$ for monodisperse, amorphous polymer melts.

The deformation mechanism maps of Gilbert et al. [3] and of Ahmad et al. [4] relied upon limited measurements in the glass transition and rubbery regimes. Moreover, they needed to calibrate their models to test data for various polymer grades with, for instance, different molecular weights and weight distributions. This motivates the current paper: our aim is to construct experimentally validated deformation maps in terms of the modulus E and the flow strength $\sigma_y$ as a function of temperature T and strain rate $\dot{\varepsilon}$ of a single commercial PMMA grade close to the glass transition. A failure envelope for the tensile failure strain as a function of temperature and strain rate is also obtained.

The present study is structured as follows. First, the deformation mechanism regimes and associated constitutive laws of Figure 1 are reviewed for a linear, amorphous polymer such as PMMA in terms of the underlying molecular deformation processes. Tensile tests on a commercial grade of PMMA are then detailed. Finally, the constitutive laws are calibrated from the measured stress-strain curves and deformation and failure maps are constructed.

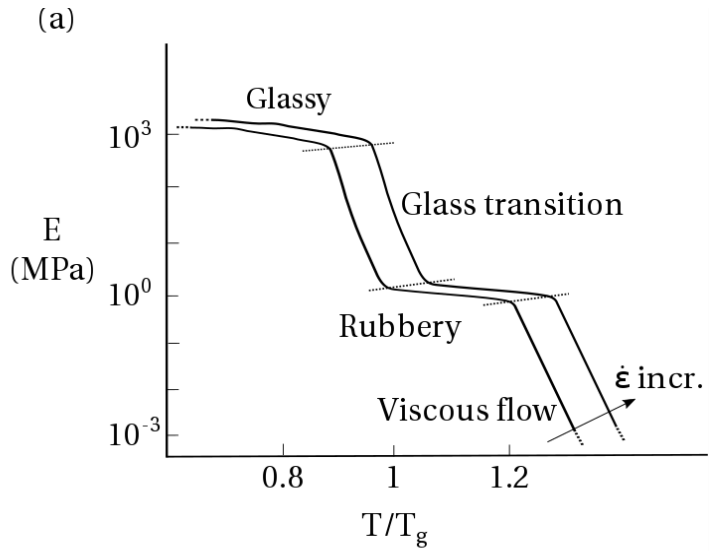

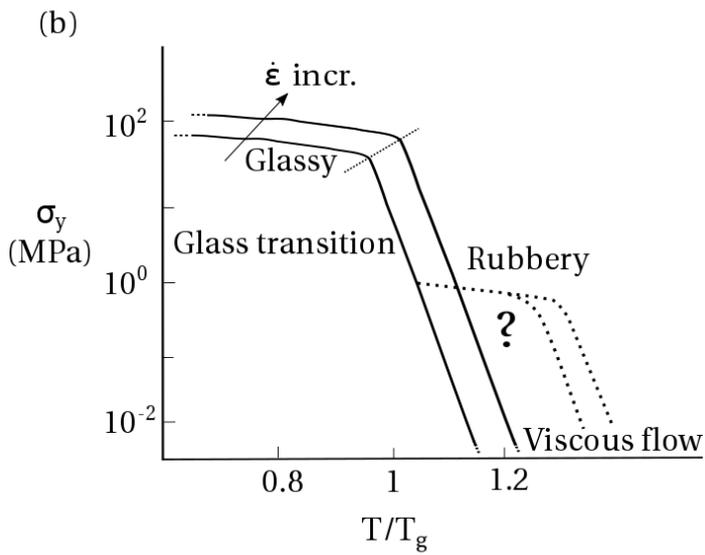

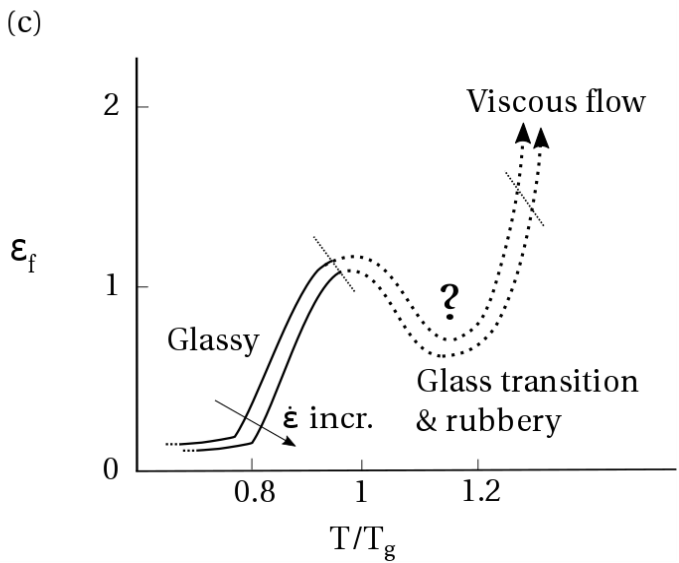

Figure 1 - Constitutive trends from the literature for a linear, amorphous polymer in uniaxial tension at temperatures close to its glass transition: (a) vanishing strain modulus as a function of $T/T_g$ for two distinct strain rates [3], (b) flow strength as a function of $T/T_g$ for two distinct strain rates [4], and (c) true tensile failure strain as a function of $T/T_g$ for two distinct strain rates [5–7].

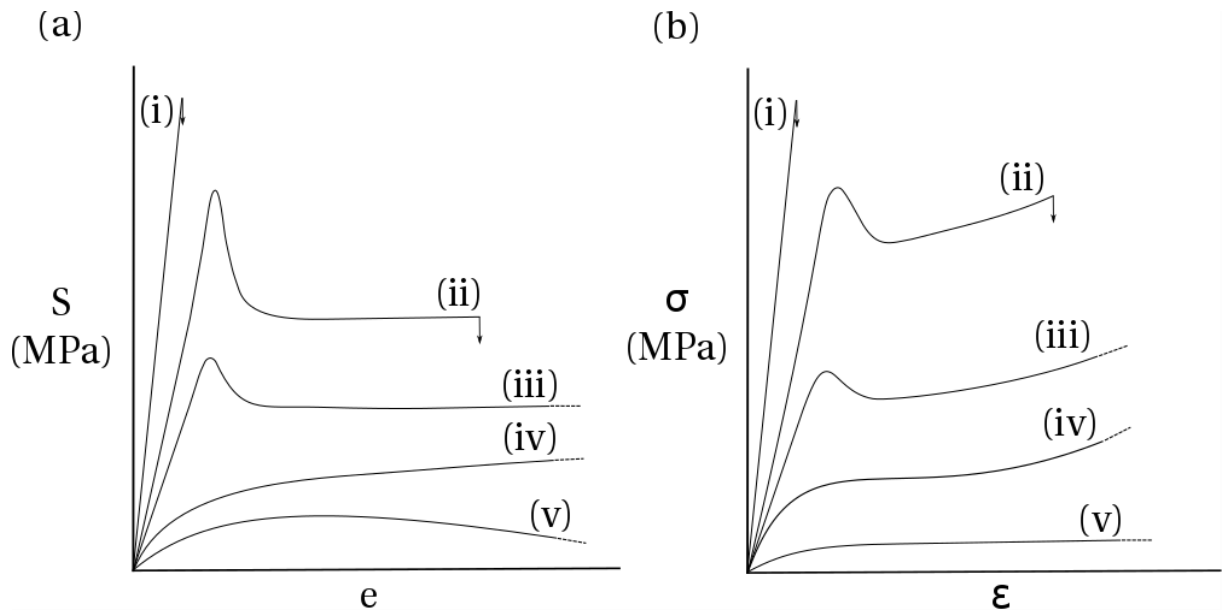

Figure 2 – Illustration of the typical stress versus strain response of PMMA in uniaxial tension, around and above the glass transition temperature for identical strain rates: (a) nominal stress versus nominal strain at constant nominal strain rate and (b) true stress versus true strain at constant true strain rate. As the temperature increases for a fixed rate, the response evolves through the following constitutive regimes: (i) the brittle (elastic) glassy, (ii) the plastic yielding glassy, (iii) the glass transition, (iv) the rubbery, and (v) the viscous flow regime.

## 2. A brief review of deformation mechanisms

Linear, amorphous polymers (such as PMMA, polycarbonate or polystyrene) exhibit a range of deformation mechanisms, depending upon temperature (relative to $T_g$) and strain rate. These include the glassy, glass transition, rubbery, viscous flow and decomposition regime in order of increasing temperature (or decreasing strain rate). For each regime (excluding decomposition), constitutive laws can be stated for the modulus E (at vanishing strain[1]), tensile flow strength $\sigma_y$ and tensile failure strain $\varepsilon_f$. We consider each regime in terms of molecular deformation mechanism and state constitutive laws for E, $\sigma_y$, and $\varepsilon_f$. Later in the paper, the measured tensile stress-strain response of the PMMA is used to calibrate these constitutive equations.

*2.1 The glassy regime and the glass transition regime*

*The small strain, viscoelastic response*
The glassy regime exists at temperatures below the glass transition, $0 < T/T_g < 1$. Following Yannas and Luise [8] and Gilbert et al. [3], we shall assume that the glassy modulus $E_g$ decreases in an almost linear fashion with increasing temperature due to the increase in vibrational energy of chain segments to give [3,8]:

$$E_g = E_0 \left(1 - \alpha_m \frac{T}{T_g}\right) \tag{1}$$

in terms of a coefficient $\alpha_m$ and modulus $E_0$ at absolute zero.

Near the glass transition temperature ($T/T_g \approx 1$) the small strain response is viscoelastic in nature and there is choice in the degree of sophistication with which to model the viscoelastic behaviour. For example, Gilbert et al. [3] treated the solid as a parallel array of a large number of one dimensional Maxwell units; each unit comprises a dashpot and a spring in series, such that the array has a spectrum of relaxation time to mimic the macroscopic, observed relaxation in modulus around $T_g$. Here, we idealise the small strain viscoelastic response of PMMA in the glassy and glass transition regime by a single Maxwell unit comprising a linear spring of glassy modulus $E_g$, as defined by Eq. (1), in series with a linear dashpot with a viscosity $\eta_{WLF}$ following WLF theory [9]:

$$\eta_{WLF} = 3\eta_0 \exp\left(\frac{-\ln(10)C_1(T/T_g - 1)}{C_2 + T/T_g - 1}\right) \quad (2)$$

where $\eta_0$ is a reference viscosity at $T_g$, and $C_1$ and $C_2$ are fitting constants. $C_1$ can be related to the free volume fraction of the polymer at $T_g$, while $C_2$ can be related to the free volume expansion coefficient of the polymer [10]. Williams et al. [9] demonstrated that the values of $C_1$ and $C_2$ are independent of the choice of polymer system (with $C_1 = 17.44$, $C_2 = 51.6$ K), although more recent studies on amorphous polymers report tailored values for $C_1$ and $C_2$, see for instance Ferry [11] and Mark [12]. The magnitude of the reference viscosity $\eta_0$ is mainly governed by the molecular weight of the polymer as discussed below.

---

[1] Note that we do not focus on the small strain harmonic viscoelastic response in this study; rather we focus on the tensile stress versus strain response at strain levels up to the failure strain.

For a uniaxial tensile test at constant strain rate $\dot{\varepsilon}$ for time $t > 0$, we shall make use of the secant modulus $E_s = \sigma(\varepsilon_{ref})/\varepsilon_{ref}$, where $\varepsilon_{ref}$ is a reference strain, taken to equal 0.05 in the present study. For the Maxwell unit under consideration $E_s$ reads as:

$$E_s = \frac{\eta_{WLF}\dot{\varepsilon}}{\varepsilon_{ref}}\left[1 - \exp\left(\frac{-\varepsilon_{ref}}{\dot{\varepsilon}}\frac{E_g}{\eta_{WLF}}\right)\right] \quad (3)$$

where $E_g$ and $\eta_{WLF}$ are defined in Eq. (1) and Eq. (2), respectively.

*The large strain, viscoplastic response*

The dependence of the plastic response in the glassy regime upon temperature is more complex, and can be sub-divided into two regimes, $0 < T/T_g < 0.8$, and $0.8 < T/T_g < 1$. Typically, when a linear, amorphous polymer is subjected to uniaxial tension at $T/T_g < 0.8$, the stress versus strain response is almost linear and brittle fracture intervenes (at a true strain below 0.1) by the formation of an unstable craze or by fast fracture from a pre-existing flaw [12,18]. Deformation occurs by the stretching and bending of secondary (van der Waals) bonds, resulting in a low sensitivity to both strain rate and temperature by secondary relaxation mechanisms [3].

For $0.8 < T/T_g < 1$, shear yielding follows an initial elastic response. Typically, yield is accompanied by a load drop and by the development of a neck. High strain rate sensitivity reduces the degree of neck formation and orientation hardening leads to neck propagation along the length of the specimen [13–15]. After the neck has propagated along the entire length of the sample, the load increases again until failure intervenes.

Near the glass transition temperature ($T/T_g \approx 1$), the van der Waals bonds melt and polymer chain segments can reptate past each other [3]. The stress versus strain response is highly rate and temperature sensitive at temperatures near $T_g$. At temperatures above $T_g$, there ceases to be a load drop at yield, and no neck is formed.

It is widely accepted that plastic yielding of linear, amorphous polymers is governed by the thermally activated motion of molecular chain segments [16,17]. The time and temperature dependence of the flow strength may be described by a modified version of the Eyring model which was originally developed for the viscous flow of liquids [18–21]. The Ree-Eyring model assumes that an applied stress $\bar{\sigma}$ (with a strict definition to be made precise below) reduces the energy barrier for vibrating chain segments to jump forward and increases the barrier to jump backwards. Between the secondary $\beta$ transition and the glass transition temperature $T_g$, the von Mises, effective plastic strain rate $\dot{\varepsilon}_e = (2\dot{\varepsilon}_{ij}^p \dot{\varepsilon}_{ij}^p / 3)^{1/2}$ is related to $\bar{\sigma}$ via a single transition process such that [17]:

$$\frac{\dot{\varepsilon}_e}{\dot{\varepsilon}_0} = \sinh\left(\frac{\bar{\sigma}v}{kT}\right)\exp\left(\frac{-q}{kT}\right) \qquad (4)$$

where $\dot{\varepsilon}_0$ is a reference strain rate, q is an activation energy, v is an activation volume, and k is Boltzmann's constant. Linear, amorphous polymers are pressure-dependent in their flow response, such that yield is activated by both the von Mises effective stress $\sigma_e$ and the hydrostatic (mean) stress $\sigma_h$, where[2]:

$$\sigma_e^{\;2} = \frac{3}{2} S_{ij} S_{ij} \qquad (5)$$

and

$$\sigma_h = \frac{\sigma_{kk}}{3} \qquad (6)$$

---

[2] A repeated suffix denotes summation over 1 to 3 following the usual Einstein convention.

in terms of the deviatioric stress tensor $S_{ij}$ $(= \sigma_{ij} - \delta_{ij}\sigma_h)$. Note that $\sigma_h$ is equal in magnitude but opposite in sign to the pressure p. A common assumption is to assume that the stress measure $\bar{\sigma}$, which activates plastic flow, is given by:

$$\bar{\sigma} = \sigma_e + \alpha\sigma_h \qquad (7)$$

where $\alpha$ is a pressure sensitivity index taken to be a material constant, see for example Ward [22]. Thus, the tensile yield strength $\sigma_{ty} > 0$ (at a prescribed value of strain rate $\dot{\varepsilon}_e$) reads:

$$\sigma_{ty} = \sigma_e = \bar{\sigma}(1 + \alpha/3)^{-1} \qquad (8)$$

while the compressive yield strength $\sigma_{cy} < 0$ (at the same value of strain rate $\dot{\varepsilon}_e$) reads:

$$\sigma_{cy} = -\sigma_e = -\bar{\sigma}(1 - \alpha/3)^{-1} \qquad (9)$$

Consequently, the ratio $|\sigma_{cy}|/\sigma_{ty}$ is:

$$\frac{|\sigma_{cy}|}{\sigma_{ty}} = \frac{3 + \alpha}{3 - \alpha} \qquad (10)$$

For PMMA, measurements of this ratio imply $\alpha \approx 0.4$ [4,20]. The consensus of experimental evidence suggests that plastic flow of PMMA is almost incompressible [23]. Consequently, for the isotropic case we can assume that:

$$\dot{\varepsilon}_{ij}^p = \frac{3}{2}\frac{S_{ij}}{\sigma_e}\dot{\varepsilon}_e \qquad (11)$$

We shall assume that Eq. (4) describes the large strain viscoplastic response of PMMA in the glassy regime and in the glass transition regime.

*Failure strain*

Within the plastic yielding regime, $0.8 < T/T_g < 1$, large plastic strains are achievable prior to ductile fracture. Failure is by the stable growth of cavities (known as 'diamonds') originating from surface defects or crazes until they reach a critical size [24]. A few experimental studies report the brittle and ductile tensile failure strain of PMMA as a function of temperature and strain rate in the glassy regime [5,16]. To a first approximation, the dependence of $\varepsilon_f$ upon

T/$T_g$ is governed by two linear relations, one for brittle elastic behaviour (small slope) and one for failure at large plastic strains (large slope), see Figure 1c. The experimental data of Cheng et al. [5] suggest that ductility is relatively insensitive to strain rate in the glassy regime.

*2.2 The rubbery regime*

*The small strain, viscoelastic response*
Linear, amorphous polymers of moderate to high molecular weight typically exhibit a rubbery plateau at a temperature regime just above the glass transition; the extent of the rubbery plateau depends on the molecular weight (distribution) of the polymer [25]. Within this rubbery regime, most of the secondary bonds are broken and long range sliding of polymer chains is prevented by chain entanglements. The entanglement points act as physical cross-links governing the polymer's constitutive state in a similar fashion to chemical cross-links of rubbers. Instead of making use of entropy-based models that relate the rubbery modulus to temperature (see Treloar et al. [26]), we find that the following empirical relation is adequate to describe the rate and temperature sensitivity of the modulus in the rubbery regime:

$$E = E_R^0 \left(1 - \alpha_R \frac{T}{T_g}\right) \left(\frac{\dot{\varepsilon}}{\dot{\varepsilon}_R}\right)^n \qquad (12)$$

where $\alpha_R$ is a temperature coefficient, n is a rate sensitivity index and $E_R^0$ and $\dot{\varepsilon}_R$ refer to a reference modulus and strain rate, respectively.

*The large strain, viscoplastic response*
Ahmad et al. [4] omit the presence of the rubbery plateau between the glassy and viscous flow regimes in their description of $\sigma_y$ as a function of rate and temperature. As the width of the rubbery plateau is governed by the average molecular weight (distribution), this approach may suffice for grades of low molecular weight. In contrast, commercial linear amorphous polymers typically possess a high molecular weight, and an extensive rubbery regime on a $\sigma_y$ versus T/$T_g$ plot is anticipated above the glass transition. Here, we assume a linear elastic stress-strain behaviour to represent the rubbery response of PMMA. For the flow strength versus temperature plot it is necessary to introduce a reference strain $\varepsilon_{ref}$ such that the flow 'strength' is given by:

$$\sigma_y = E_R \varepsilon_{ref} \qquad (13)$$

with $E_R$ governed by Eq. (12).

*Failure strain*

In contrast to the abundance of rate and temperature dependent constitutive relations for E and $\sigma_y$, data on the effect of rate and temperature on the tensile failure strain $\varepsilon_f$ of linear, amorphous polymers in the glass transition and rubbery regime are scarce. Smith [6] investigated the ultimate tensile strength and failure strain of cross-linked amorphous elastomers such as styrene-butadiene rubbers (SBR) near their glass transition and demonstrated the applicability of time-temperature superposition for the tensile failure strain. Now consider the ductility in the vicinity of the rubber to viscous flow transition. Vinogradov et al. [7] found that the tensile failure strain $\varepsilon_f$ of polymer melts went through a minimum at T far above $T_g$ and then underwent a steep increase as the viscous flow regime is approached. The dependence of $\varepsilon_f$ upon $T/T_g$ for a commercial linear, amorphous polymer may exhibit a similar trend: a local maximum is attained in the rubbery regime, followed by a dip as the polymer transitions from the rubbery to the viscous state, see Figure 1c. However, insufficient measurements are reported in the literature for firm conclusions to be drawn. For instance, tensile tests on some commercial grades of PMMA and polystyrene melts show that the failure strain is independent of strain rate [27]. Fundamental models on the dependence of failure strain upon molecular weight (distribution) in the rubbery state also appear to be lacking [28,29].

*2.3 The viscous flow regime*

*The small strain, viscoelastic response*

At temperatures somewhat above the glass transition temperature ($T/T_g > 1.1$), molecular entanglement points slip and the macroscopic constitutive behaviour can be treated as a viscous melt. Flow occurs by the reptation of individual polymer chains that are confined in tubes as defined by their surrounding environment [30–32]. Simple models exist for this behaviour. For example, Gilbert et al. [3] modelled the rate and temperature dependent small-strain modulus in the viscous regime by means of a Maxwell unit consisting of a WLF or Arrhenius governed dashpot and a linear spring with a temperature dependent modulus.

*The large strain, viscoplastic response*

Ahmad et al. [4] suggest a straightforward view of the rate and temperature dependent flow strength $\sigma_y$ for a polymer melt in the viscous flow regime. They assume that the flow strength $\sigma_y$ versus strain rate $\dot{\varepsilon}$ relation is linear with slope $3\eta$, where $\eta$ is the shear viscosity. They assert that the temperature dependence of $\eta$ is given by WLF theory, see Eq. (2), for temperatures close to $T_g$, and by an Arrhenius relation for higher temperatures [4]. Doi and Edwards [32–34] developed a reptation model for viscous flow, and thereby showed that the shear viscosity has a power law dependence upon molecular weight, with an exponent of 3 in good agreement with experimental observations for polymer melts [11,35].

It is broadly accepted that the rupture strain of a viscous melt is governed by the surface tension-driven instability of Rayleigh-Plateau [36,37].

*2.4 More sophisticated three dimensional constitutive models*

Boyce and co-workers [14,38–41] have developed a sophisticated three dimensional theoretical framework to simulate the stress-strain response of linear, amorphous polymers over the entire glass-to-rubber transition regime. They assume that the polymer has a parallel combination of intermolecular and molecular network resistances. The Boyce et al. [42] framework was originally developed for PETG, but more recently Palm et al. [43] have used it to describe the constitutive behaviour of PMMA around $T_g$ for uniaxial compression. The three dimensional models of G'Sell and Souahi [44] and Dooling et al. [45] focus on the constitutive response of PMMA loaded in uniaxial tension around and above its glass transition. These three-dimensional large strain constitutive models give good curve fits to uniaxial compressive and tensile test data. However, they make use of a large number of fitting constants and there are few non-proportional, multi-axial tests available to calibrate the full three-dimensional response. Moreover, these models were developed to predict the deformation response, but give little insight into failure.

# 3. Materials and methods

*3.1 Test material*

Cast PMMA sheets (t = 3 mm) were acquired from Altuglas International (Arkema, France). The $T_g$ (= 116°C) was measured by Differential Scanning Calorimetry (DSC) with a heating rate of 10°C min$^{-1}$. The molecular weight distribution of the samples was measured via Gel Permeation Chromatography[3] (GPC) with tetrahydrofuran as the eluent and detection by refractive index. The molecular weight distribution is shown in . The weight-average molecular weight $M_w$ equals 3 580 000 g mol$^{-1}$, while the number-average molecular weight $M_n$ equals 607 000 g mol$^{-1}$, resulting in a polydispersity index $M_w/M_n$ equal to 5.9.

*3.2 Specimen geometries and testing procedures*

Dogbone specimens were machined from the PMMA sheets as shown in Figure 4, and were adhered[4] to aluminum alloy tabs (35 mm x 26 mm x 1.4 mm). The specimens were pin-loaded and dots of diameter equal to 1 mm were painted onto the gauge section of the specimens with white acrylic paint. Their relative displacement was tracked by a video camera during each test. The average nominal strain e in the gauge section of the specimen was measured from the axial separation between the two pairs of horizontally aligned white dots. Assuming uniform deformation in the gauge section, the true strain $\varepsilon$ is given by:

$$\varepsilon = \ln(1 + e) \tag{14}$$

The dogbone specimens were tested in a servo-hydraulic test machine in displacement control, and the test temperature was set by making use of an environmental chamber with feedback temperature control and fan-assisted air circulation. The specimen's temperature was independently monitored by a placing a thermocouple against it. The front face of the dogbone specimens was viewed through a transparent window of the environmental chamber. The cross-head displacement, tensile load P, air chamber temperature, and sample temperature were simultaneously monitored and logged by a PC. Assuming uniform and incompressible deformation in the gauge section, the true stress is computed by:

---

[3] GPC was performed with an Agilent Technologies PL GPC220 (USA) with a nominal flow rate equal to $1.67 \cdot 10^{-5}$ l/s at a testing temperature equal to 30°C.
[4] Tabs were glued with Loctite EA 9497 high temperature resistant structural bonding epoxy (Henkel, Germany). The adhesive was allowed to cure at room temperature for at least 48 hours.

$$\sigma = \frac{P}{A} = \frac{P}{A_0}(1+e) \tag{15}$$

where A and $A_0$ refer to the current and nominal cross-sectional area of the gauge, respectively.

Tests were performed for temperatures ranging from 80 °C to 185 °C and for three cross-head velocities ($v_{ch}$ = 5 mm s$^{-1}$, $v_{ch}$ = 0.5 mm s$^{-1}$ and $v_{ch}$ = 0.05 mm s$^{-1}$). The tests were terminated at the point of specimen failure or when the maximum actuator displacement (100 mm) was reached. The latter constraint corresponds to a true strain of approximately unity within the gauge section of the dogbone specimens. Upon recognizing that higher failure strains occur when the test temperature is at or above the glass transition temperature, additional tests were performed using an alternative hourglass-shaped specimen used to probe the failure strain at temperatures near $T_g$ (in the range of 115 °C to 190 °C) following the practice of Hope et al. [46].

*3.3 Strain measurements for the hourglass specimens*

Consider the geometry of the hourglass specimen as shown in Figure 5. Recall that the loading direction is aligned with the 3-direction, the thickness direction with the 2-direction, and the transverse direction is aligned with the 1-direction. The axial strain can be measured by the change in shape of the dots at mid-section (longitudinal measurement method). However, this method is prone to scatter, and the alternative transverse gauge-based measurement method of Hope et al. [46] was used in order to reduce this measurement scatter:

(i) An average measure for the transverse true tensile strain $\varepsilon_1$ at the minimum section of hourglass specimen is obtained from the spacing w between the pair of white marker dots.

(ii) Assume that the through-thickness strain $\varepsilon_2$ equals the transverse strain $\varepsilon_1$. This was verified by post-failure examination of the thickness and width of the hourglass specimens.

(iii) Assume incompressibility, as discussed below, such that $\varepsilon_1 + \varepsilon_2 + \varepsilon_3 = 0$.

Then, the true tensile strain is related to the current spacing w and initial spacing $w_0$ by:

$$\varepsilon_3 = -2\ln\left(\frac{w}{w_0}\right) \tag{16}$$

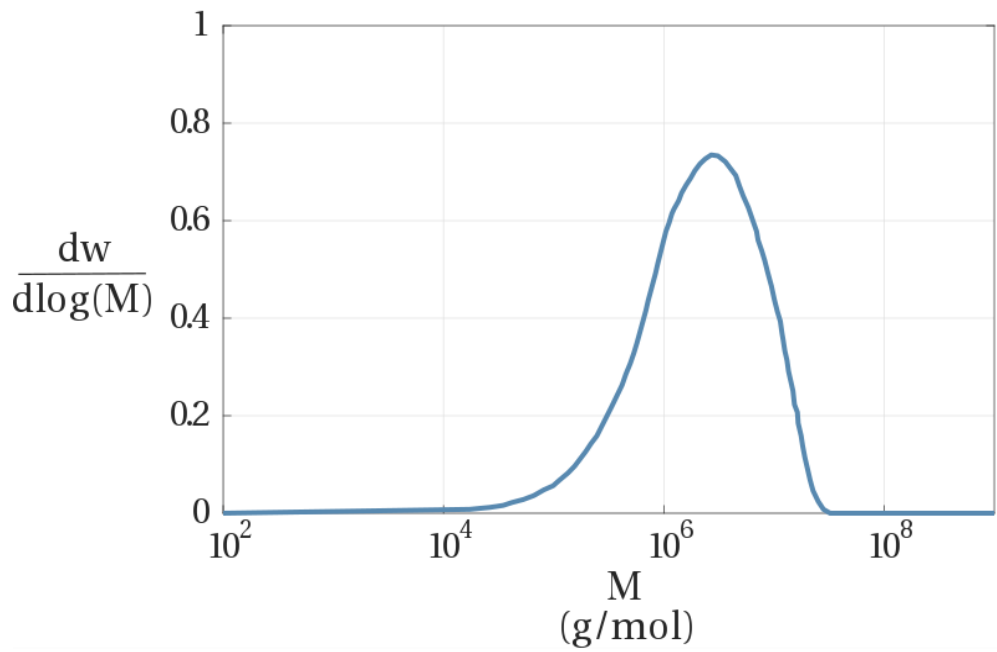

Figure 3 – Measured molecular weight distribution of the cast PMMA sheet by Gel Permeation Chromatography. M and w refer to the molecular weight and molecular weight fraction, respectively.

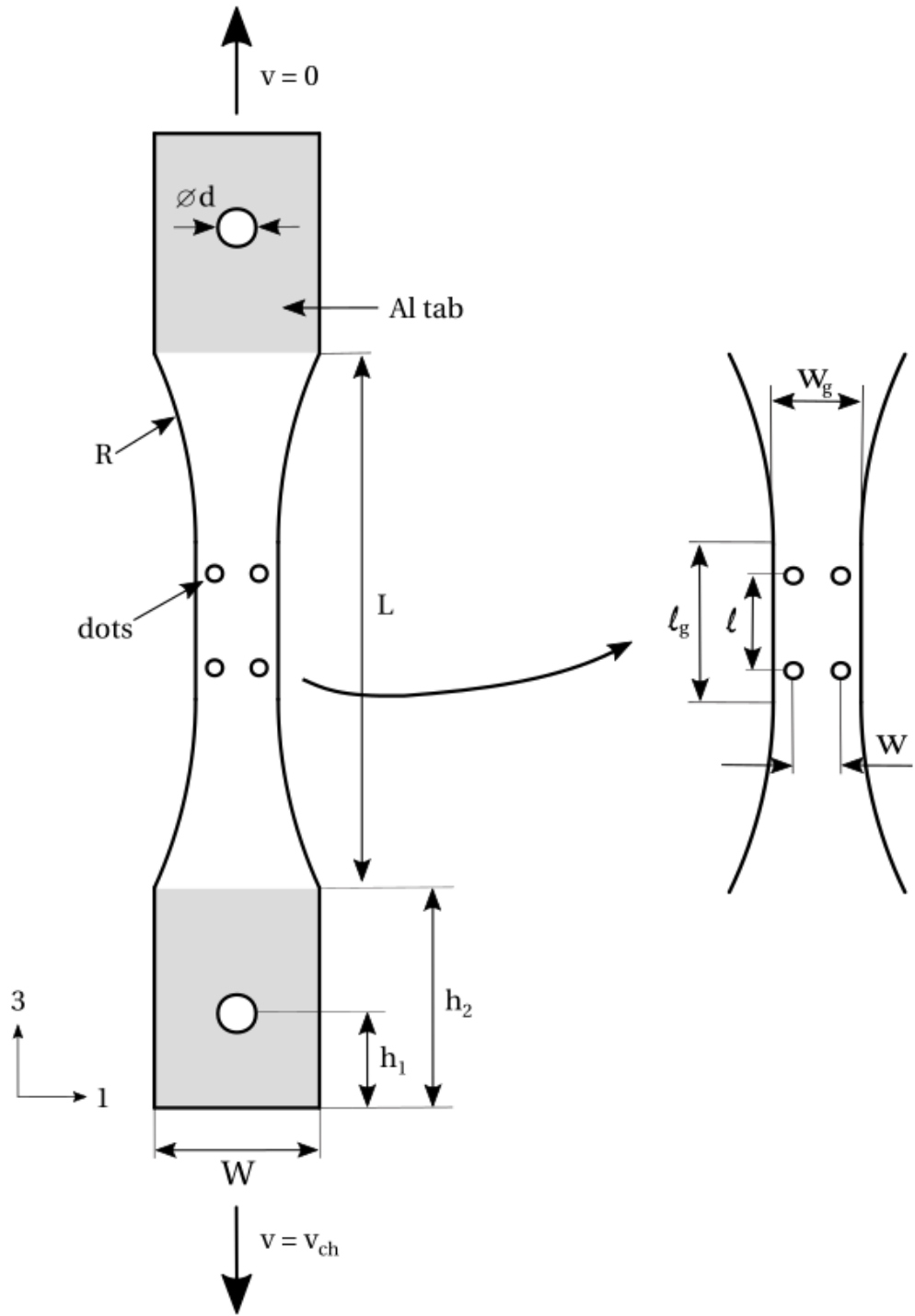

Figure 2 – Plan view of the dogbone tensile specimen deformed at a cross-head velocity $v_{ch}$ with nominal thickness equal to 3mm, R = 72.5 mm, L = 85 m, d = 6 mm, W = 26 mm, $h_1$ = 15 mm, $h_2$ = 35 mm. Nominal dimensions in the gauge area: $w_g$ = 13 mm, $l_g$ = 25 mm, w = 10 mm, l = 15 mm.

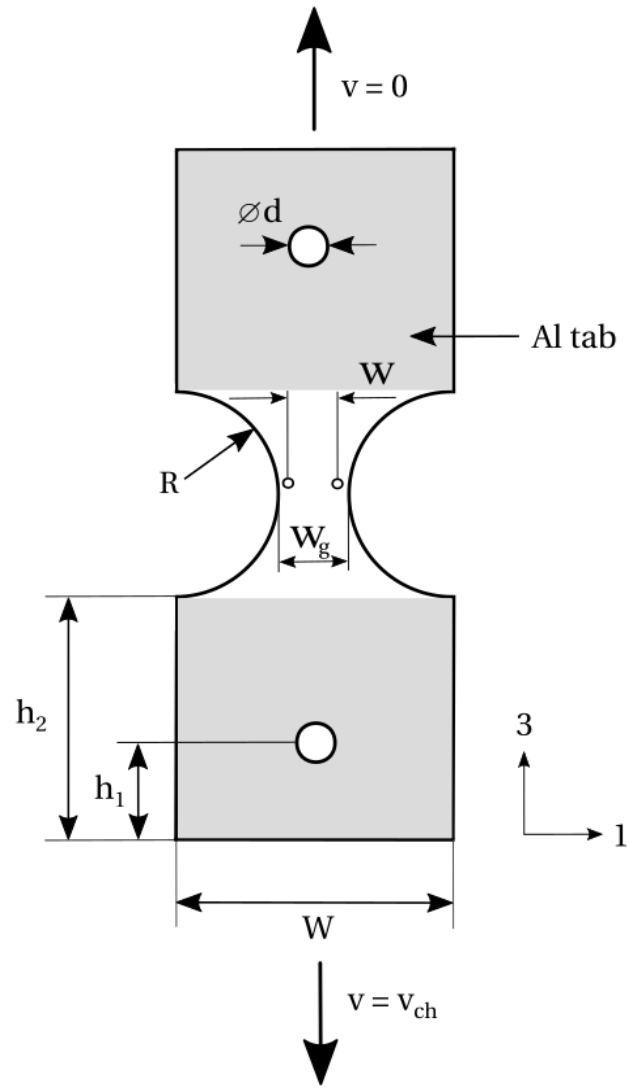

Figure 3 – Plan view of the hourglass-shaped tensile specimen deformed at a cross-head velocity $v_{ch}$ with nominal thickness equal to 3 mm, R = 15 mm, d = 6 mm, $h_1$ = 15 mm, $h_2$ = 35 mm, W = 40 mm. Nominal dimensions in the gauge area: $w_g$ = 10 mm, w = 9 mm.

# 4. Results and discussion

True tensile stress versus true tensile strain curves are reported for a temperature T in the range 80°C to 185°C, corresponding to $T/T_g$ in the range 0.91 to 1.18, and nominal strain rates in the range $5.9 \times 10^{-4}$ s$^{-1}$ to $5.9 \times 10^{-2}$ s$^{-1}$. Values for the modulus, flow strength and failure strain are extracted from each of the curves and used to calibrate the constitutive descriptions of the previous section. A series of tests were also performed for temperatures in the range of 20°C and 80°C in order to identify the temperature (and strain rate) at which the behavior switches from elastic-brittle to elastic-plastic. The transition occurs at close to 70°C, such that the response is elastic-brittle at the high strain rate and elastic-plastic at the low strain rate (and intermediate rate). We focus on the elastic-plastic response of the PMMA in this study.

The transverse true strain was measured as a function of axial true strain for all dogbone specimen tensile tests. The elastic Poisson's ratio $\nu$ was found to increase from $\nu = 0.35$ at T = 80°C to $\nu = 0.4$ at T = 105°C, in agreement with the measurements by Lu et al. [47] and Gilmour et al. [48]. In contrast, the plastic Poisson's ratio $\nu_p$, defined as the ratio of the transverse to axial value of the true plastic strain, was almost constant at 0.45 – 0.5 for 80°C ≤ T ≤ 185°C, implying that plastic flow is almost incompressible [23].

*4.1 Stress-strain curves*

The true stress versus true strain responses of the dogbone specimens are plotted in Figure 6a for $0.91 \leq T/T_g \leq 1.01$ and in Figure 6b for $1.04 \leq T/T_g \leq 1.18$ for a nominal strain rate of $5.9 \times 10^{-3}$ s$^{-1}$.

At $T < T_g$, linear, elastic deformation takes place prior to yielding. The yield point is followed by material softening, and by the associated development of a neck. The severity of necking is assessed by measuring by the ratio r of the gauge width in the necked zone over the gauge width in the unnecked zone. The value of r at a true strain equal to 0.55 is shown in Figure 6a for $0.91 \leq T/T_g \leq 1.01$. The degree of necking decreases as $T/T_g$ increases and necking ceases around the glass transition. Note that the true stress versus true strain responses as shown in Figure 6a for $0.91 \leq T/T_g \leq 0.97$ give the overall structural response in the presence of necking rather than the local material response. We shall make later use of the modulus and initial peak strength, and not discuss further the neck development. See, for example, Wu and van der Giessen [49] for an analysis of necking of polymers in the presence of strain softening.

Now consider the response at higher temperatures, in the range of T = 120°C to T = 155°C, (corresponding to $1.01 \leq T/T_g \leq 1.1$). The tensile response in this regime is almost independent of temperature and strain rate, and this is characteristic of rubberlike elasticity, see Figure 6b. The temperature and rate sensitivity increase again at the highest testing temperatures (T = 170°C and T = 185°C, i.e. $1.14 \leq T/T_g \leq 1.18$), indicating that a transition occurs to a viscous deformation regime.

Additional insight is obtained by interrupting selected tensile tests prior to failure, and then by fully unloading them. The loading-unloading stress-strain curves are shown in Figure 7. For $T/T_g = 0.94$ (within the glassy regime), elastic unloading occurs in the manner of an elasto-viscoplastic solid, with a substantial remnant strain at zero load. When the temperature is increased to $T/T_g = 1.06$, the elastic rubbery regime is entered and the unloading curve is almost coincidental with the loading curve; marginal hysteresis occurs and the remnant strain upon unloading is close to zero. Finally, at $T/T_g = 1.16$, the viscous regime is just entered and unloading is accompanied by remnant strain.

*4.2 Modulus as a function of rate and temperature*

The dependence of the modulus E upon $T/T_g$ is given in Figure 8a for three nominal strain rates. Note that E is measured as the secant modulus based on a true reference strain $\varepsilon_{ref}$ equal to 0.05. For comparison, the small deformation storage modulus is included, as obtained by the Dynamic Mechanical Analysis (DMA) of a single cantilever beam at an excitation frequency equal to 0.1Hz and heating rate of 5°C/min. The modulus, as measured by DMA, is in relatively close agreement to the measured values in a tensile test at a strain rate of $5.9 \times 10^{-2}$ s$^{-1}$.

The dependence of E upon $T/T_g$ and upon strain rate in the glassy and glass transition regime is captured by means of Eq. (3). Equation (3) is fitted by non-linear regression with the Matlab Curve Fitting Toolbox for the measured modulus values corresponding to $0.91 \leq T/T_g \leq 1.04$ for $\varepsilon_{ref} = 0.05$. The fitted parameters ($\eta_0$, $C_1$, $C_1$, $E_0$, $\alpha_m$) are stated in . The dependence of E upon $T/T_g$ in the rubbery regime is assumed to be governed by Eq. (12). Equation (12) is calibrated by the measured modulus values corresponding to $1.06 \leq T/T_g \leq$ 1.18, resulting in $E_R^0 = 65.8$ MPa, $\alpha_R = 0.80$, $\dot{\varepsilon}_R = 1.58$ s$^{-1}$ and $n = 0.173$. The E versus T/T$_g$ curves, as predicted by the fitted versions of Eq. (3) and Eq. (12) for the three tested nominal strain rates, are depicted in Figure 8a. The experimental trends are captured to adequate accuracy.

*4.3 Flow strength as a function of rate and temperature*

The dependence of the measured flow strength[5] $\sigma_y$ upon T/T$_g$ for three nominal strain rates is shown in Figure 8b. The flow strength in the glassy and glass transition regime is assumed to be governed by rate and temperature via a single transition process Ree-Eyring equation, see Eq. (4). Equation (4) is fitted by non-linear regression to the $\sigma_y$ values corresponding to $0.91 \leq T/T_g \leq 1.04$ by aid of the Matlab Curve Fitting Toolbox. The fitted values for v, $\dot{\varepsilon}_0$, and q as well as the initial values for the regression (taken from Bauwens-Crowet [20]) are summarized in . The fitted value for the activation volume is in the same order of magnitude as the ones reported in the literature [22]. This volume is about 5 to 10 times larger than the estimated volume of a monomer unit which is in agreement with the idea that shear yielding is governed by the collective movement of numerous chain segments [16]. As depicted Figure 8b, the measured $\sigma_y$ versus T/T$_g$ trends in the glassy and glass transition regime are well approximated by the Ree-Eyring fit, see Eq. (4), while the dependence of the flow strength in the rubbery regime is captured with reasonable accuracy by Eq. (13) with $\varepsilon_{ref} = 0.05$ and the fitted version of Eq. (12) for the rubbery modulus $E_R$.

*4.4 Failure strain as a function of rate and temperature*

The measured true tensile failure strain $\varepsilon_f$ is plotted[6] as a function of T/T$_g$, for two nominal strain rates $\dot{e} = 5.9 \times 10^{-2}$ s$^{-1}$ and $\dot{e} = 5.9 \times 10^{-4}$ s$^{-1}$, in Figure 9a. Recall that the failure strain is measured via a transverse gauge in addition to a longitudinal gauge over the full range of T/T$_g$ employed. The two methods are in good agreement except at T/T$_g \geq 1.1$ for which $\varepsilon_f \geq 1.5$. In this regime, the reduction in cross-sectional area at the minimum section of the hourglass-shaped specimen is larger than the elongation of the dots. We conclude that the transverse gauge gives a more accurate value for the local failure strain.

---

[5] In the glassy regime, the tensile flow strength $\sigma_y$ ($= \sigma_{ty}$) is defined as the value of the true tensile stress corresponding to the peak in load [22]. This distinct yield point disappears upon entering the glass transition regime. The flow strength in the glass transition and rubbery regime is therefore defined as the value of the true stress at a reference true strain equal to 0.05.

[6] The specimen did not fail at T = 185 °C at the maximum actuator displacement for $\dot{e} = 5.9 \times 10^{-2}$ s$^{-1}$.

The transverse gauge-based failure envelope for $\dot{e} = 5.9 \times 10^{-4}$ s$^{-1}$ is fitted by a 3$^{rd}$ order polynomial ($R^2 = 0.97$):

$$\varepsilon_f = -80.8 \left(\frac{T}{T_g}\right)^3 + 230.2 \left(\frac{T}{T_g}\right)^2 - 211 \left(\frac{T}{T_g}\right) + 62.9 \tag{17}$$

while the transverse gauge-based failure envelope for $\dot{e} = 5.9 \times 10^{-2}$ s$^{-1}$ is fitted by a linear relation ($R^2 = 0.92$):

$$\varepsilon_f = 7.3 \frac{T}{T_g} - 6.3 \tag{18}$$

Equations (17) and (18) are plotted in Figure 9a along with the observed values of failure strain.

The residual true tensile strain $\varepsilon_r$, measured post-failure, is compared with $\varepsilon_f$ in Figure 9b. At temperatures $T/T_g < 1$, there is negligible elastic spring-back and $\varepsilon_r \approx \varepsilon_f$. For $1 \leq T/T_g < 1.1$, $\varepsilon_r$ decreases sharply with increasing temperature, consistent with rubbery behavior. For $1.1 \leq T/T_g < 1.18$, both $\varepsilon_r$ and $\varepsilon_f$ increase with increasing temperature, consistent with a transition to viscous behaviour.

The curve fits (Eq. (17) and Eq. (18)) from the measured values of $\varepsilon_f$ as a function of $T/T_g$ are compared in Figure 9c with the observed failure strains as reported by Cheng et al. [5] for PMMA in the glassy regime. There is good agreement between the data of Cheng et al. [5] and those of the present study. We also make the comparison of PMMA in the rubbery regime to that of a lightly cross-linked elastomer styrene-butadiene rubber (SBR), as reported by Smith [50]. The failure strain envelope above $T_g$ of the linear, amorphous, high molecular weight PMMA is found to be similar to the failure strain envelope above $T_g$ of a chemically cross-linked SBR.

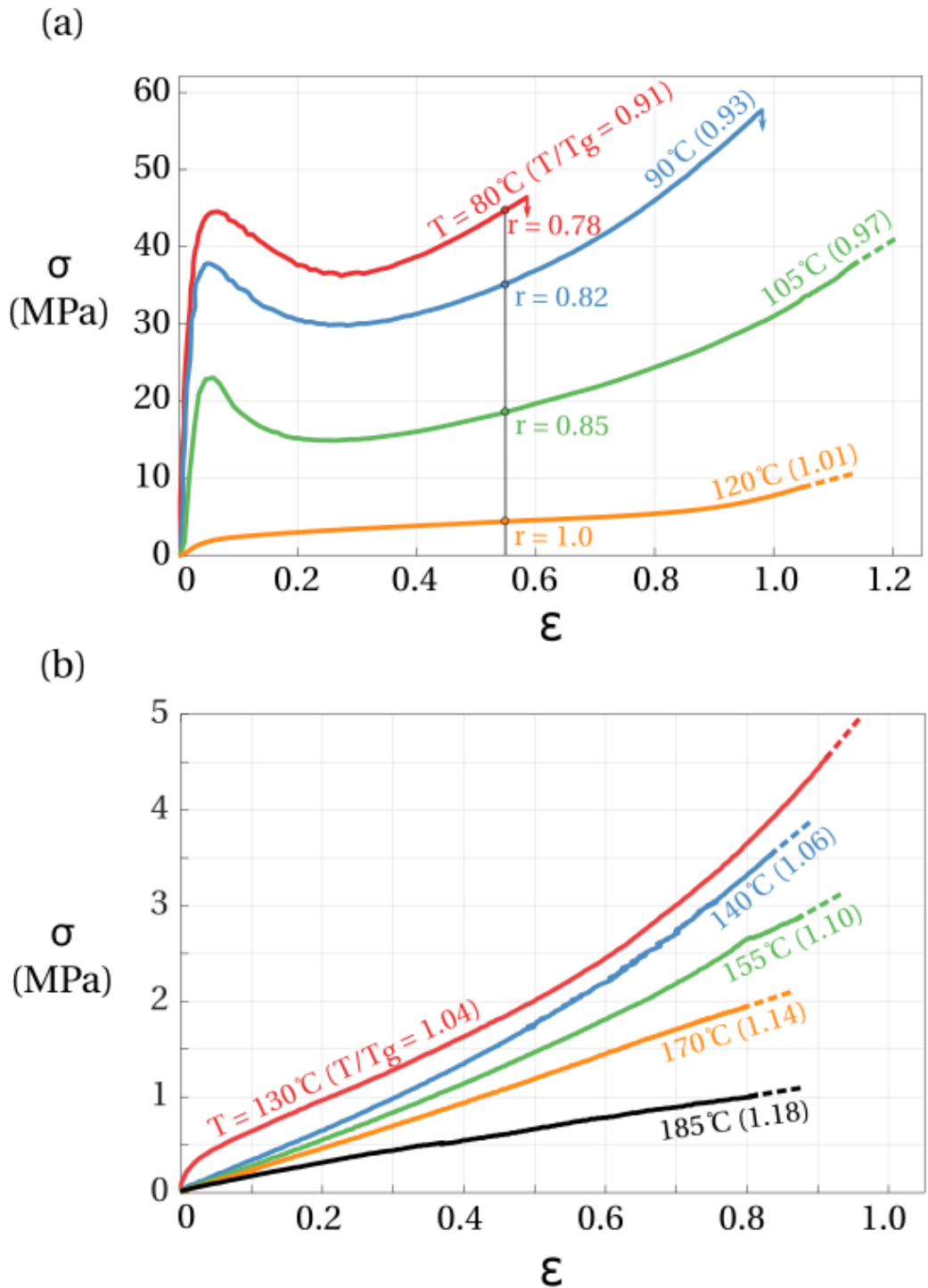

Figure 4 – Measured true stress versus true strain curves for a nominal strain rate $\dot{e}$ = 5.9 x10$^{-3}$ s$^{-1}$ and testing temperatures ranging from (a) T = 80 ℃ to T = 120 ℃ and (b) T = 130 ℃ to T =185 ℃. A downward arrow at the end of the curve denotes sample failure, while a dashed line refers to the end of test when the maximum displacement of the actuator is attained. The r value in (a) represents the ratio of the measured gauge width in the necked zone over the gauge width in the unnecked zone at a true strain equal to 0.55.

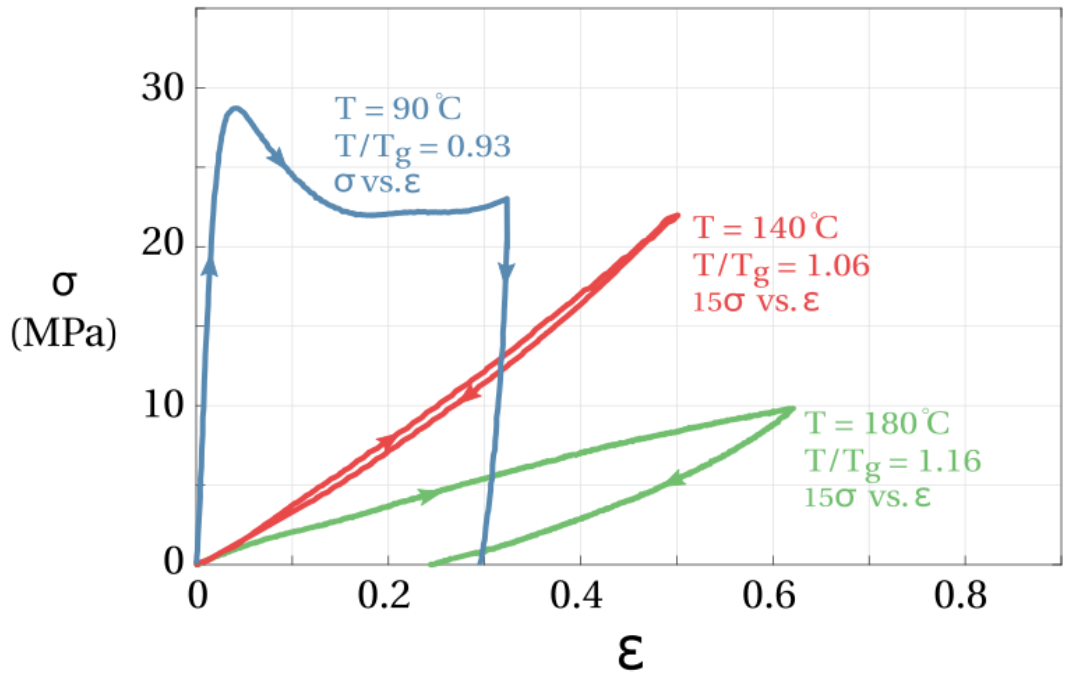

Figure 5 – Loading/unloading true stress versus true strain curves for three different testing temperatures though identical cross-head loading and unloading velocities ($\dot{e} = 5.9 \times 10^{-4}$ s$^{-1}$).

Table 1 – Summary of the fitted and initial Eq. (3) parameters to the measured secant modulus values corresponding to 80 °C ≤ T ≤ 130 °C via nonlinear regression in Matlab, $R^2 = 0.97$ for $\varepsilon_{ref} = 0.05$.

|  | Fitted value | Gilbert et al. [3] |
|---|---|---|
| $\eta_0$ (Pa·s) | $6.7 \cdot 10^8$ | $1.1 \cdot 10^{16}$ |
| $C_1$ | 42.77 | 17.4 |
| $C_2$ (K) | 113.8 | 143 |
| $E_0$ (MPa) | 3522 | 8600 |
| $\alpha_m$ | 0.85 | 0.28 |

Table 2 – Summary of the fitted and initial Eq. (4) parameters for the measured flow strength values corresponding to 80 °C ≤ T ≤ 130 °C via nonlinear regression in Matlab, $R^2 = 0.98$.

|  | Fitted value | Bauwens-Crowet et al. [20] |
|---|---|---|
| v (nm$^3$) | 1.62 | 2.56 |
| $\dot{\varepsilon}_0$ (s$^{-1}$) | $1.5 \cdot 10^{56}$ | $2 \cdot 10^{56}$ |
| q (J) | $7.31 \cdot 10^{-19}$ | $6.85 \cdot 10^{-19}$ |

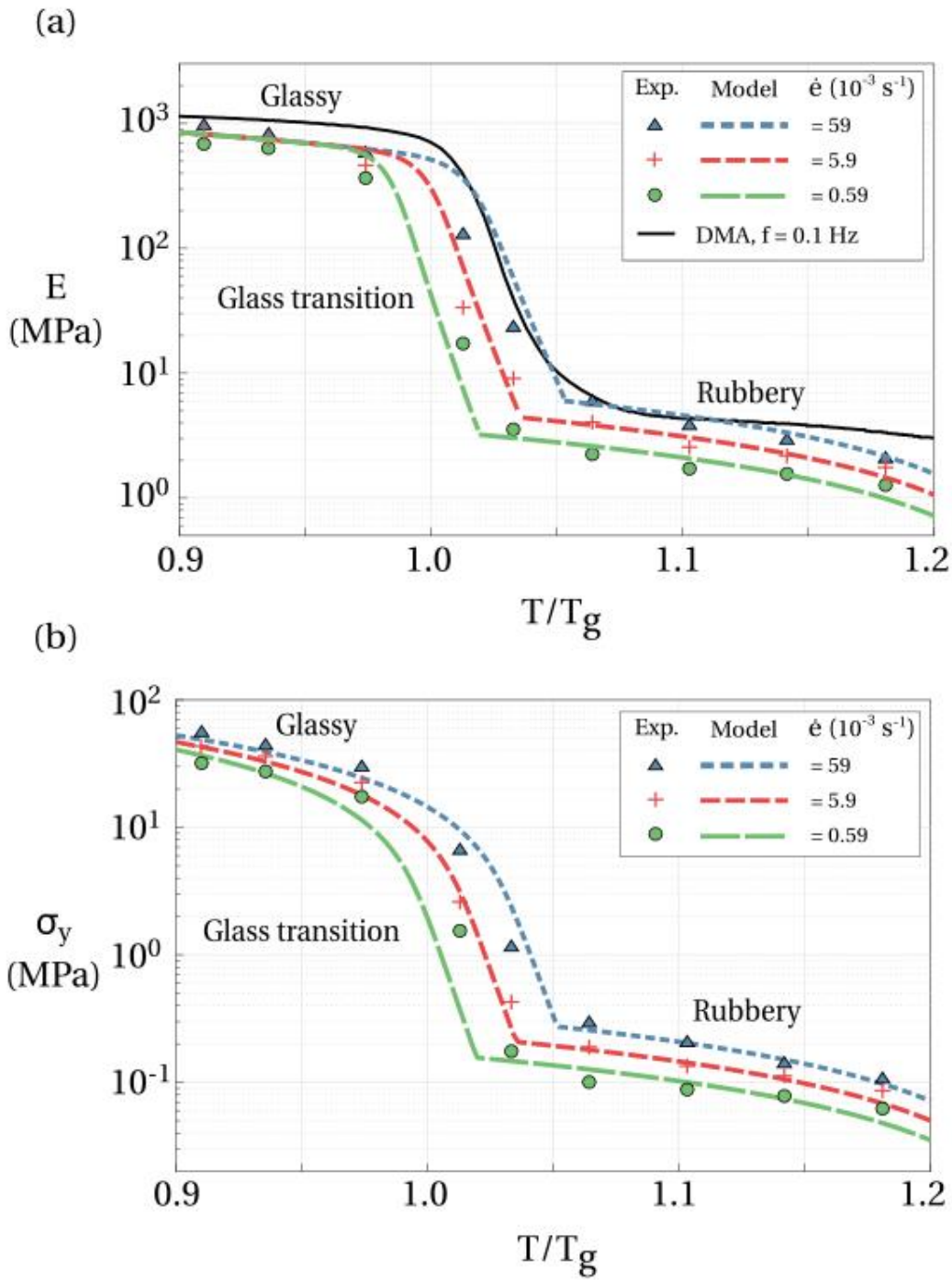

Figure 8 - Deformation diagrams of the tested PMMA grade: (a) measured secant modulus versus $T/T_g$ along with the E - $T/T_g$ relations predicted via the fitted versions of Eq. (3) and Eq. (12), (b) measured flow strength $\sigma_y$ versus $T/T_g$ along with the $\sigma_y$ - $T/T_g$ relations predicted via the fitted versions of Eq. (4) and Eq. (13) for $\varepsilon_f = 0.05$.

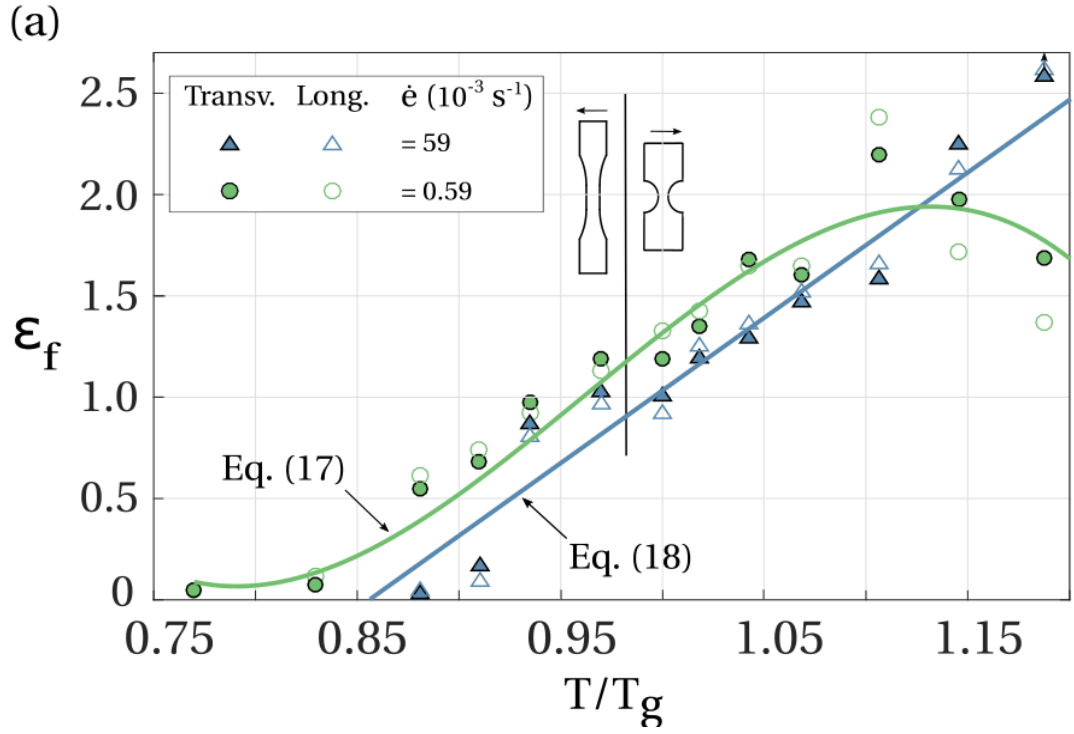
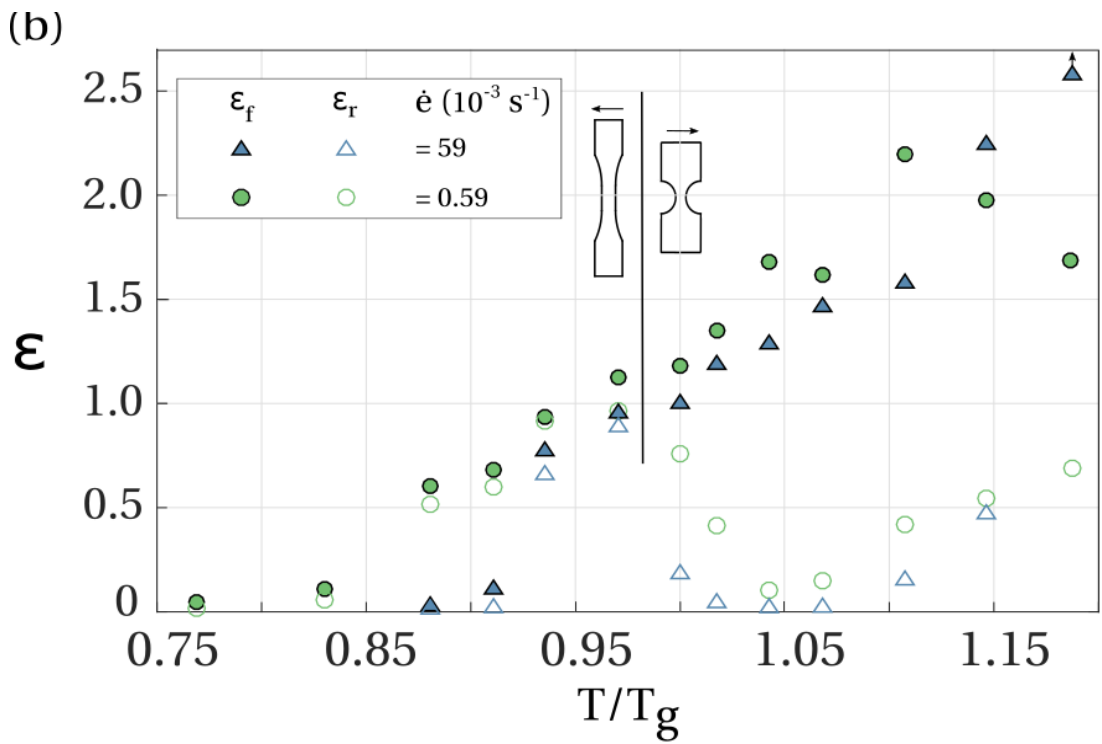

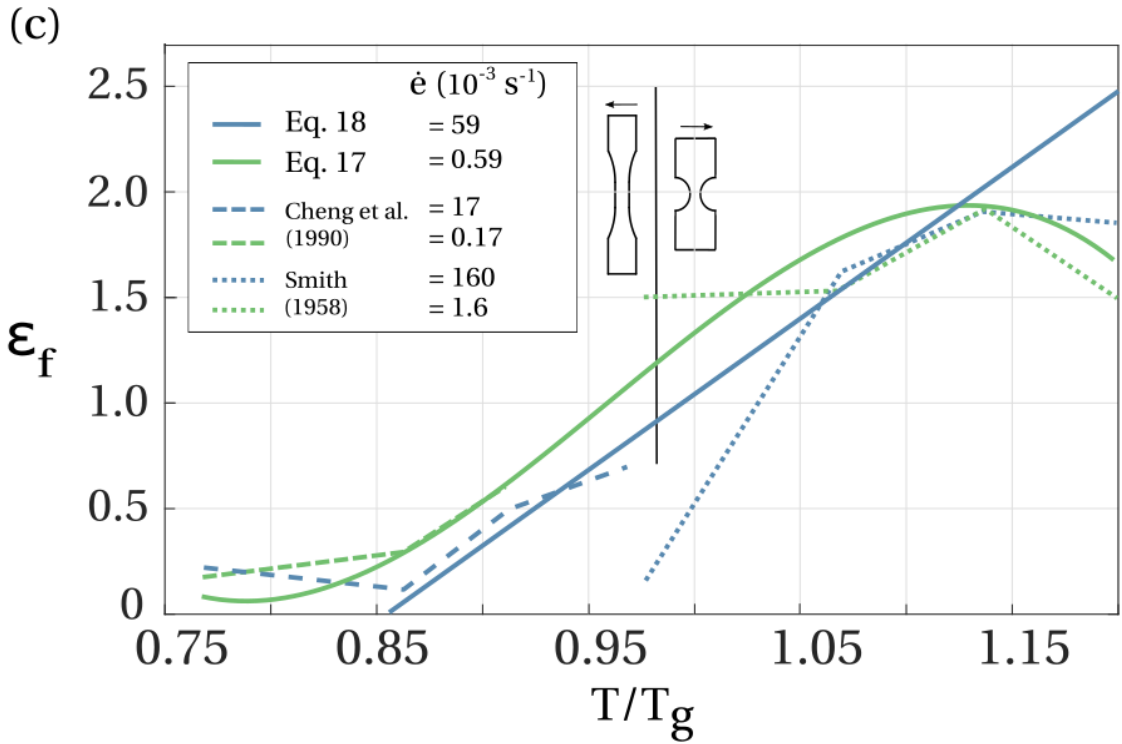

Figure 9 - PMMA failure diagrams: (a) true tensile failure strain $\varepsilon_f$ versus $T/T_g$ for nominal strain rates $\dot{e} = 5.9\times10^{-4}$ s$^{-1}$ (circle) and $\dot{e} = 5.9\times10^{-2}$ s$^{-1}$ (triangle). Closed symbols refer to transverse strain gauge measurements (transv.), while open symbols represent longitudinal gauge measurements (long.). The solid lines depict the fitted $\varepsilon_f$ versus $T/T_g$ relations (Eq. (17) and Eq. (18)). (b) residual true tensil strain $\varepsilon_r$ (open symbol) versus $T/T_g$ and tensile failure strain $\varepsilon_f$ (closed symbol) versus $T/T_g$. Note that in (b) the strain values for the dogbone specimens are based on the longitudinal gauge measurement method, while the strain values for the the hourglass-shaped specimens are based on the transverse measurement method. (c) Fitted $\varepsilon_f$ versus $T/T_g$ relations (Eq. (17) and Eq. (18)) along with the reported experimental $\varepsilon_f$ versus $T/T_g$ trends by Cheng et al. [5] for PMMA in the glassy regime and Smith [50] for SBR in the glass transition and rubbery regime.

## 5. Concluding remarks

The tensile response of a commercial polymethyl methacrylate (PMMA) grade has been characterized over a range of temperatures near the glass transition and over two decades of strain rate via a series of uniaxial tensile tests. Modulus, flow strength and failure strain are plotted as a function of temperature via deformation and failure maps for selected strain rates. Fitted constitutive relations in terms of modulus and flow strength are reported for three identified constitutive regimes: the glassy, glass transition, and rubbery regime. The models can be generalized in a straightforward fashion to three dimensional behaviour assuming isotropy. It is emphasized that the constitutive descriptions (including failure) are needed in order to model the processing of PMMA close to the glass transition temperature. Failure is of prime importance in processes ranging from melt blow moulding to solid-state foaming.

## Acknowledgements

This work was supported by the Engineering and Physical Sciences Research Council, award number 1611305. The authors are also grateful for the financial support from SABIC and the technical assistance from Dr. Martin van Es. We also acknowledge financial support from the ERC MULTILAT, grant number 669764.